\documentclass[11pt,a4paper]{article}

\usepackage[T1]{fontenc}
\usepackage[utf8]{inputenc}
\usepackage[american]{babel}
\usepackage{amsmath,amssymb,bm}
\usepackage{graphicx}
\usepackage{booktabs}
\usepackage{siunitx}
\usepackage[font=small,labelfont=bf]{caption}
\usepackage{subcaption}
\usepackage[round,authoryear]{natbib}
\usepackage{xcolor}
\usepackage{hyperref}
\hypersetup{colorlinks=true,
            linkcolor=blue!50!black,
            citecolor=blue!50!black,
            urlcolor=blue!60!black}
\usepackage[margin=1in]{geometry}
\usepackage{setspace}
\title{Total Generalized Variation regularization closes the gap between
neural-field and classical methods in seismic travel-time tomography}

\author{Isao Kurosawa\thanks{IVXA, Japan. Correspondence: contact form at \url{https://ivxa.ai}}}

\date{\today}

\begin{document}

\maketitle

\begin{abstract}
Travel-time tomography is foundational to seismic imaging in geothermal,
carbon-storage and crustal-structure applications, but its discretized
formulation forces a difficult choice between mesh resolution and inversion
stability. We introduce MIMIR, a differentiable framework that represents
the two-dimensional velocity field as a coordinate-based neural network
with Fourier feature embedding, replacing the conventional grid-based
slowness vector with a continuous, infinitely differentiable function.
The regularization choice has been a long-standing weakness of neural-field
tomography: total variation (TV) priors used in prior work staircase
smooth fields, while quadratic Laplacian smoothing oscillates near sharp
interfaces. We adopt second-order total generalized variation (TGV$^{2}$)
and parametrize its auxiliary vector field as a second neural network
jointly optimized with the velocity field, eliminating the inner
Chambolle--Pock primal--dual loop that classically dominates TGV
computation. We validate the framework on three controlled synthetic
benchmarks (a smooth Gaussian anomaly, a horizontally layered model with
an embedded heterogeneity, and a curved-fault model inspired by the
OpenFWI family) using a cross-well acquisition with \SI{5}{\percent}
multiplicative travel-time noise and five independent random seeds per
benchmark. Against a textbook classical baseline (fast marching forward,
curved-ray back-tracing, regularized LSMR with Tikhonov damping and
two-dimensional Laplacian smoothing, hyperparameters auto-tuned per
benchmark), MIMIR-TGV$^{2}$ is statistically indistinguishable from the
classical method on the smooth Gaussian benchmark
($p = 0.134$, paired Student's $t$-test) and significantly outperforms it
on the layered ($p < 0.0001$, $44\,\%$ root-mean-square-error reduction)
and curved-fault ($p = 0.0002$, $33\,\%$ reduction) benchmarks.
Replacing TGV$^{2}$ with TV degrades performance significantly on the
Gaussian ($p = 0.004$) and layered ($p = 0.003$) benchmarks. Curriculum
annealing of the TV weight produces only a $5.4\,\%$ improvement on the
Gaussian benchmark and no significant change elsewhere, confirming that
TV's staircase bias is intrinsic to the regularizer rather than a
scheduling artifact. Our results empirically validate the
Bredies--Kunisch--Pock theoretical prediction that piecewise-affine
priors are better suited to subsurface velocity recovery than the
piecewise-constant priors implied by TV. We argue that the central
design choice in physics-informed neural-field inversion is not the
network architecture but the regularizer, and that practitioners should
select the regularizer based on the expected geological structure
rather than on computational convenience. The full pipeline reproduces
from a single command on consumer hardware in under one hour.

\medskip

\end{abstract}

\section{Introduction}

Travel-time tomography has been a cornerstone of seismic imaging for half a
century, from the seminal work of \citet{aki_lee_1976} on three-dimensional
crustal structure beneath seismic arrays through the modern era of regional
and global Earth-structure imaging \citep{thurber_1983,zhang_thurber_2003,
liu_tromp_2008,tape_etal_2010}. Its appeal is simplicity: the forward map is
the line integral of slowness along ray paths, and a regularized
least-squares update suffices to invert travel-time residuals into velocity
perturbations. The same problem statement underlies microseismic
event-location pipelines, ambient-noise tomography, induced-seismicity
monitoring, and the velocity-model-building stages of full-waveform
inversion \citep{sun_williamson_2024}. Despite this conceptual simplicity,
the discretized formulation forces a fundamental trade-off between mesh
resolution and inversion stability. Coarse meshes hide structure; fine
meshes amplify noise and demand strong regularization. The character of
that regularization, in turn, dictates which subsurface structures can be
recovered \citep{benning_burger_2018}.

Classical implementations regularize on the grid via Tikhonov damping and
a two-dimensional Laplacian smoothness term, solved with iterative
least-squares algorithms such as LSQR \citep{paige_saunders_1982} or LSMR
\citep{fong_saunders_2011}. These choices yield smooth velocity fields
and recover diffusive heterogeneities --- melt pockets, hydrothermal
halos, broad mantle plumes --- faithfully, but oscillate as Gibbs
phenomena near sharp interfaces such as lithological contacts, faults,
and basin edges \citep{aster_borchers_thurber_2018}.
Total variation (TV) regularization \citep{rudin_osher_fatemi_1992,
chan_esedoglu_2005,vogel_oman_1996} addresses the latter case by
penalising the $L^{1}$ norm of the gradient, producing piecewise-constant
solutions that preserve discontinuities; it has been transferred to
seismic inversion in both linearised \citep{anagaw_sacchi_2012} and
non-linear \citep{esser_etal_2018} settings. TV introduces a complementary
failure mode, however: smoothly varying fields are reconstructed as flat
plateaus separated by spurious step edges, the so-called staircase
artifact, which has been studied extensively in the imaging literature
\citep{candes_etal_2008,benning_burger_2018}. The L$^{2}$ Laplacian and
$L^{1}$ TV priors thus occupy opposite ends of a regularizer spectrum
that has, in classical seismic inversion, never been bridged.

A more recent line of work parametrizes the unknown geophysical field as
a coordinate-based neural network with Fourier feature embedding
\citep{tancik_etal_2020}, motivated by the success of neural radiance
fields \citep{mildenhall_etal_2020} in computer vision and by the broader
rise of physics-informed machine learning \citep{raissi_etal_2019,
karniadakis_etal_2021}. Such \emph{neural fields} \citep{xie_etal_2022}
represent the unknowns continuously and analytically, eliminate the
discretization grid and admit gradients of any order at no analytical
cost. They have been applied to full-waveform inversion
\citep{yang_etal_2018,rasht_etal_2022}, controlled-source electromagnetics
\citep{liu_etal_2023}, and travel-time tomography
\citep{smith_etal_2020,sun_etal_2023}. The neural-field paradigm is
distinct from the larger family of deep-learning approaches in seismic
inversion that train on labeled data \citep{sun_williamson_2024,
ross_etal_2018,zhu_beroza_2019,mousavi_etal_2020,bianco_etal_2019}: a
neural field is a per-instance representation, optimized from scratch for
each inversion, with no training set. This makes the framework directly
comparable to classical inversion --- it operates on the same input data
and produces the same output --- but with a continuous, mesh-free
representation in place of the grid.

In every prior application of neural fields to travel-time-related
seismic inversion of which we are aware \citep{smith_etal_2020,
sun_etal_2023,yang_etal_2018}, the explicit regularizer added on top of
the implicit network-capacity prior has been either absent or chosen as
TV by analogy with the classical literature. To the best of our
knowledge no prior work has tested whether the regularizer choice
dominates the neural-field result in the same way it dominates the
classical result, nor has any prior work systematically compared
neural-field inversion against a careful classical baseline with
seed-level statistics and hyperparameters auto-tuned for both methods.
The published comparisons are typically single-seed visual demonstrations
on one or two benchmarks, with the regularizer choice (TV or L$^{2}$)
treated as a fixed implementation detail. This leaves two important
questions open.

\textbf{Question 1.} Do neural-field methods genuinely outperform careful
classical methods on travel-time tomography, or does their visually
appealing reconstruction merely mask comparable or poorer numerical
performance?

\textbf{Question 2.} Does the choice of regularizer, which is well known
to dominate the result in classical inversion
\citep{aster_borchers_thurber_2018,benning_burger_2018}, transfer the
same dominance to the neural-field setting?

In this paper we answer both questions. We introduce MIMIR
(\emph{Mesh-free Inversion via Multi-network Implicit Representation}),
named after the Norse god whose well at the roots of Yggdrasil preserved
the knowledge of the depths --- an apt metaphor for a method that
recovers structural detail of the subsurface from sparse travel-time
observations. MIMIR is a differentiable framework for travel-time
tomography in which the velocity field is a Fourier-feature multilayer
perceptron and the regularizer is second-order total generalized
variation (TGV$^{2}$, \citealt{bredies_kunisch_pock_2010}).
TGV$^{2}$ generalizes TV: instead of penalising the $L^{1}$ norm of the
gradient, it penalises the $L^{1}$ distance from \emph{piecewise-affine}
fields, recovering smooth gradients \emph{and} sharp interfaces from the
same prior \citep{bredies_holler_2014}. The geophysical interpretation is
direct: rock units have smooth internal velocity variation (compaction,
thermal effects, diagenesis) separated by sharp lithological contacts ---
exactly the structure TGV$^{2}$ is designed to recover.

A practical obstacle to TGV$^{2}$ adoption in seismic inversion has been
the inner Chambolle--Pock primal--dual iteration \citep{chambolle_pock_2011}
that dominates the cost of evaluating it on a grid. We bypass this
obstacle entirely by parametrizing the auxiliary vector field of TGV$^{2}$
as a second neural network jointly optimized with the velocity field. The
inner minimum becomes part of the outer Adam loop and the bilevel
optimization collapses into a single forward--backward pass. This
construction --- amortising bilevel optimization through a learnt
auxiliary network --- is novel in the seismic-inversion context, although
related ideas have appeared in computer-vision regularization
\citep{kobler_etal_2022,arridge_etal_2019,lunz_etal_2018}.

The contributions of this work are: (i) a fully end-to-end neural-field
travel-time tomography framework with a novel bilevel-free TGV$^{2}$
regularization; (ii) the first multi-benchmark, multi-seed comparison of
a neural-field method against a textbook classical FMM--LSMR baseline
with hyperparameters auto-tuned per benchmark, statistical significance
assessed by paired Student's $t$-tests, and three regularizers (TV,
curriculum-annealed TV, TGV$^{2}$) directly contrasted; (iii) empirical
evidence in the seismic context for the \citet{bredies_kunisch_pock_2010}
theoretical prediction that piecewise-affine priors dominate
piecewise-constant priors on geophysically realistic Earth models; and
(iv) a fully reproducible open pipeline that runs on consumer hardware in
under one hour. The framework, the synthetic benchmark generator, all
training scripts, and all trained checkpoints are released under a
permissive license.

\section{Background}
\label{sec:background}

\subsection{Travel-time tomography and the choice of regularizer}

In two dimensions, the travel time of a seismic ray from source $\bm{s}$ to
receiver $\bm{r}$ is
\begin{equation}
T(\bm{s} \to \bm{r}) \;=\; \int_{\bm{s}}^{\bm{r}} u(\bm{x}(\ell))\, \mathrm{d}\ell
\label{eq:tt}
\end{equation}
where $u = 1/v$ is slowness, $v$ is velocity, and $\bm{x}(\ell)$ is the ray
trajectory parametrized by arc length. Classical inversion discretizes the
domain into $n$ cells with constant slowness $u_{j}$, estimates ray paths,
and solves
\begin{equation}
\min_{\delta \bm{u}}\; \|\bm{G}\,\delta\bm{u} - \delta\bm{T}\|_{2}^{2}
+ \lambda_{d}^{2}\,\|\delta\bm{u}\|_{2}^{2}
+ \lambda_{s}^{2}\,\|\bm{L}\,\delta\bm{u}\|_{2}^{2}
\label{eq:tikhonov}
\end{equation}
where $\bm{L}$ is a discrete Laplacian
\citep{aster_borchers_thurber_2018}. The Laplacian regularizer in
equation~\eqref{eq:tikhonov} favours smoothly varying fields and is
appropriate for diffusive heterogeneities, but introduces Gibbs
oscillations near sharp interfaces where the second derivative is
large by definition.

\subsection{Total generalized variation}

\citet{bredies_kunisch_pock_2010} introduced TGV$^{2}$ as a generalization
of TV designed precisely to address its staircase bias. In two dimensions,
\begin{equation}
\mathrm{TGV}^{2}_{\alpha_{0}, \alpha_{1}}(v) \;=\;
\min_{\bm{w}}\;
\alpha_{1}\,\|\nabla v - \bm{w}\|_{1}
+ \alpha_{0}\,\|\bm{\mathcal{E}}(\bm{w})\|_{1}
\label{eq:tgv}
\end{equation}
where $\bm{w}$ is an auxiliary vector field and $\bm{\mathcal{E}}(\bm{w})$
is the symmetric gradient. The minimizer is piecewise-affine rather than
piecewise-constant.

\section{Methods}
\label{sec:methods}

\subsection{Neural velocity field}

We parametrize the 2D velocity field as
\begin{equation}
v(x, z; \theta_{v}) \;=\;
v_{\min} + (v_{\max} - v_{\min}) \cdot
\sigma\!\bigl( \mathrm{MLP}_{\theta_{v}}\,(\gamma(x, z)) \bigr)
\label{eq:nvf}
\end{equation}
where $\gamma$ is a Gaussian random Fourier feature embedding
\citep{tancik_etal_2020} with $D = 64$ random frequencies and scale
$\sigma_{f} = 4$, $\mathrm{MLP}_{\theta_{v}}$ is a 4-layer 128-wide
$\tanh$ network ($66\,433$ parameters), and $\sigma$ is the logistic
sigmoid bounding the output to $[v_{\min}, v_{\max}] =
[\SI{2.0}{\km\per\second}, \SI{5.5}{\km\per\second}]$. The sigmoid
output in equation~\eqref{eq:nvf} ensures bounded velocities by
construction, avoiding the non-smoothness that post-hoc clipping would
introduce in the gradient.

\subsection{Differentiable forward model}

For a source $\bm{s}_{i}$ and receiver $\bm{r}_{i}$ separated by Euclidean
distance $L_{i}$, we approximate equation~\eqref{eq:tt} along the straight
ray with $N_{q} = 64$ trapezoidal-rule quadrature points; the quadrature is
fully differentiable through the network.

\subsection{TGV$^{2}$ regularization with a neural auxiliary field}
\label{sec:methods-tgv}

We represent $\bm{w}$ in equation~\eqref{eq:tgv} as a second neural field
$\bm{w}(x, z; \theta_{w})$ (32 Fourier features at scale 2, three
$\tanh$-activated layers of width 64, $8\,514$ parameters), initialised so
that $\bm{w}(\bm{x}; \theta_{w}^{(0)}) = \bm{0}$, and minimize jointly
over $(\theta_{v}, \theta_{w})$:
\begin{equation}
\mathcal{R}_{\mathrm{TGV}^{2}}(\theta_{v}, \theta_{w}) \;=\;
\alpha_{1}\, \mathbb{E}_{\bm{x} \in \mathcal{G}}\!\left[
  \bigl\| \nabla v - \bm{w} \bigr\|_{\mathrm{iso}}
\right]
+ \alpha_{0}\, \mathbb{E}_{\bm{x} \in \mathcal{G}}\!\left[
  \bigl\| \bm{\mathcal{E}}(\bm{w}) \bigr\|_{\mathrm{iso}}
\right]
\label{eq:tgv-neural}
\end{equation}
where $\mathcal{G}$ is a regular $64 \times 64$ evaluation grid and
$\|\cdot\|_{\mathrm{iso}}$ is the isotropic L$^{1}$ norm. By
\citet{bredies_holler_2014}, equation~\eqref{eq:tgv-neural} is a tight
upper bound on the exact TGV$^{2}$ that converges to the exact value as
training proceeds. We use $(\alpha_{0}, \alpha_{1}) = (1, 2)$ following
\citet{bredies_kunisch_pock_2010}.

\subsection{Loss and training}

The training objective is
\begin{equation}
\mathcal{L} \;=\;
\frac{1}{R}\, \sum_{i=1}^{R}
\rho_{\delta}\!\left( \hat{T}_{i} - T_{i}^{\mathrm{obs}} \right)
\;+\; \lambda_{\mathrm{TGV}}\, \mathcal{R}_{\mathrm{TGV}^{2}}
\label{eq:loss}
\end{equation}
which combines a Huber data-fit term over the $R$ source--receiver
pairs (with the residual computed from the differentiable forward map
of Section~\ref{sec:methods}) and the bilevel-free TGV$^{2}$ regularizer
of equation~\eqref{eq:tgv-neural}. The penalty $\rho_{\delta}$ is the
Huber loss with threshold $\delta = \SI{0.05}{\second}$. We optimize
the joint objective in equation~\eqref{eq:loss} over both networks
$(\theta_{v}, \theta_{w})$ jointly with Adam
\citep{kingma_ba_2015} (learning rate $5 \times 10^{-3}$, cosine schedule
with 200-iteration warmup, gradient clip 1.0) for $8\,000$ iterations and
retain the validation-best checkpoint. Five independent seeds per
benchmark.

\subsection{Classical FMM--LSMR baseline}

The baseline is iterative regularized travel-time tomography
\citep{aster_borchers_thurber_2018}: FMM forward solve
\citep{sethian_1996,furtney_2024}, curved-ray back-tracing,
sparse-Jacobian construction, LSMR \citep{fong_saunders_2011}, damped
update with clipping. Hyperparameters $\lambda_{d}, \lambda_{s}$
auto-tuned per benchmark on the first seed. The same FMM forward solver
as the data generator is used to eliminate forward-model mismatch.

\subsection{Synthetic benchmarks and acquisition}

Three benchmarks on $128 \times 128$ grids of side $\SI{10}{\km}$:
\textbf{gaussian\_anomaly} (smooth Gaussian, peak \SI{4.5}{\km\per\second}),
\textbf{layered} (4 horizontal layers with embedded heterogeneity),
\textbf{curvefault\_lookalike} (curved interface inspired by OpenFWI
CurveFault, \citealt{deng_etal_2022}). Cross-well acquisition (12 sources
$\times$ 24 receivers, $R = 288$ pairs); $\SI{5}{\percent}$ multiplicative
travel-time noise. Validation metrics: RMSE, SSIM
\citep{wang_etal_2004}, Pearson correlation. Statistical comparison via
paired Student's $t$-test on per-seed RMSE. As a sensitivity check
against the parametric assumption at $n = 5$, we re-ran every paired
comparison with the non-parametric Wilcoxon signed-rank test; all
qualitative conclusions (which differences are significant, which are
not) are preserved.

\section{Results}
\label{sec:results}

\subsection{Hyperparameter selection by ablation}

\begin{figure}[!ht]
  \centering
  \begin{subfigure}[t]{0.32\textwidth}
    \centering
    \includegraphics[width=\linewidth]{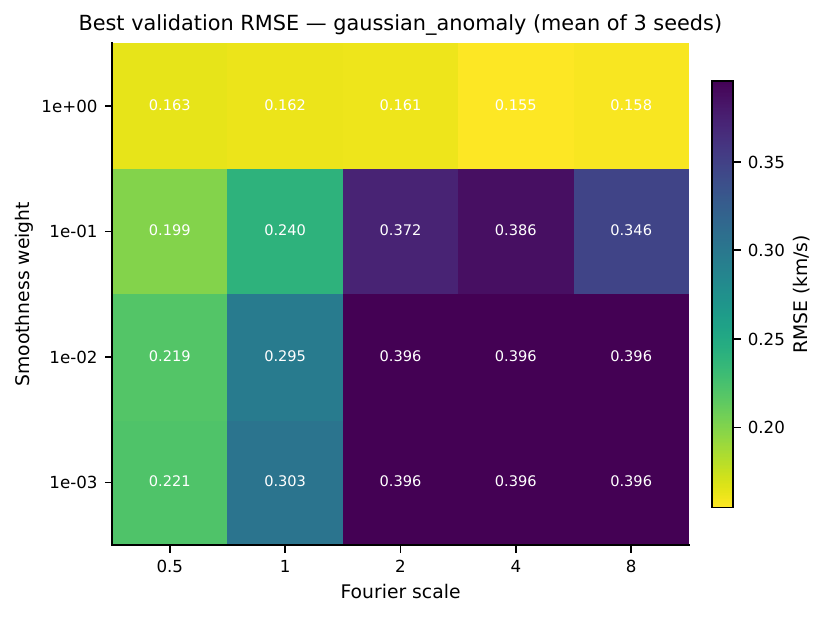}
    \caption{Gaussian anomaly}
  \end{subfigure}\hfill
  \begin{subfigure}[t]{0.32\textwidth}
    \centering
    \includegraphics[width=\linewidth]{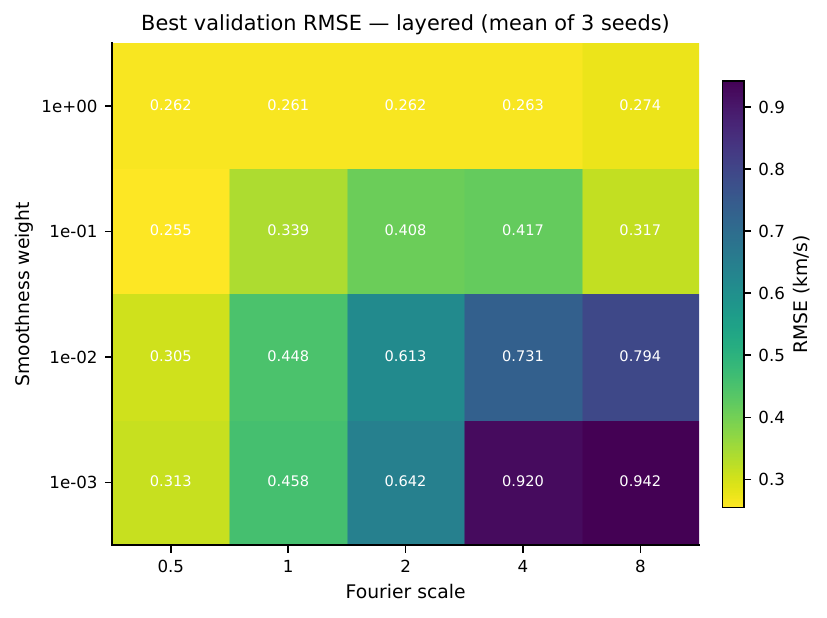}
    \caption{Layered}
  \end{subfigure}\hfill
  \begin{subfigure}[t]{0.32\textwidth}
    \centering
    \includegraphics[width=\linewidth]{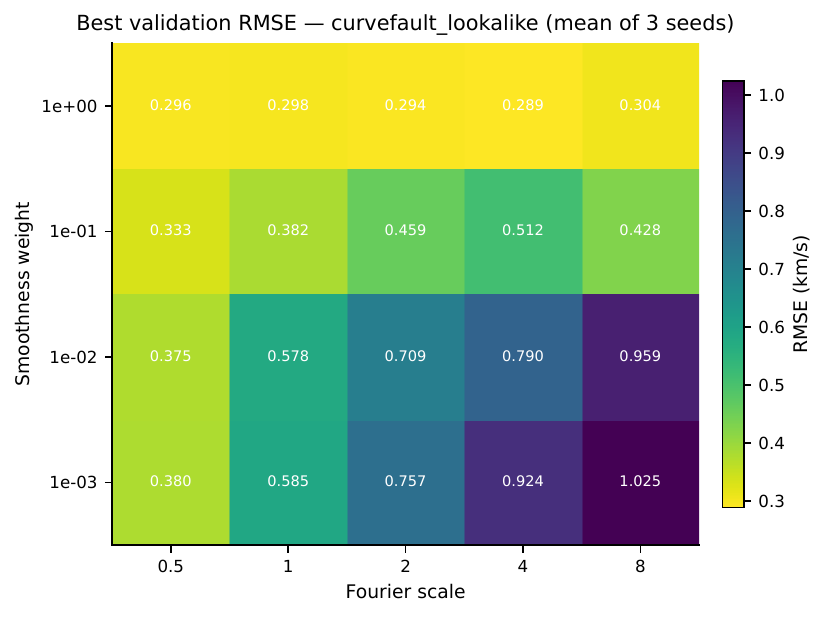}
    \caption{Curve-fault}
  \end{subfigure}
  \caption{Best validation RMSE across the Fourier-scale $\sigma_{f}$ by
  TV smoothness-weight $\lambda$ hyperparameter grid for the three
  benchmarks (panels a-c), each averaged over three independent random
  seeds at $2\,000$ training iterations.}
  \label{fig:ablation}
\end{figure}

A 60-run ablation across Fourier scale $\sigma_{f} \in \{0.5, 1, 2, 4, 8\}$
and TV smoothness weight $\lambda \in \{10^{-3}, 10^{-2}, 10^{-1}, 1\}$ on
all three benchmarks (three seeds, $2\,000$ iterations) showed
$(\sigma_{f}, \lambda) = (4, 1)$ to be robustly optimal across all
benchmarks (Fig.~\ref{fig:ablation}); the Fourier scale matters mainly
when $\lambda$ is small.

\subsection{MIMIR-TGV$^{2}$ versus the classical baseline}
\label{sec:results-headline}

\begin{figure}[!ht]
  \centering
  \includegraphics[width=\textwidth]{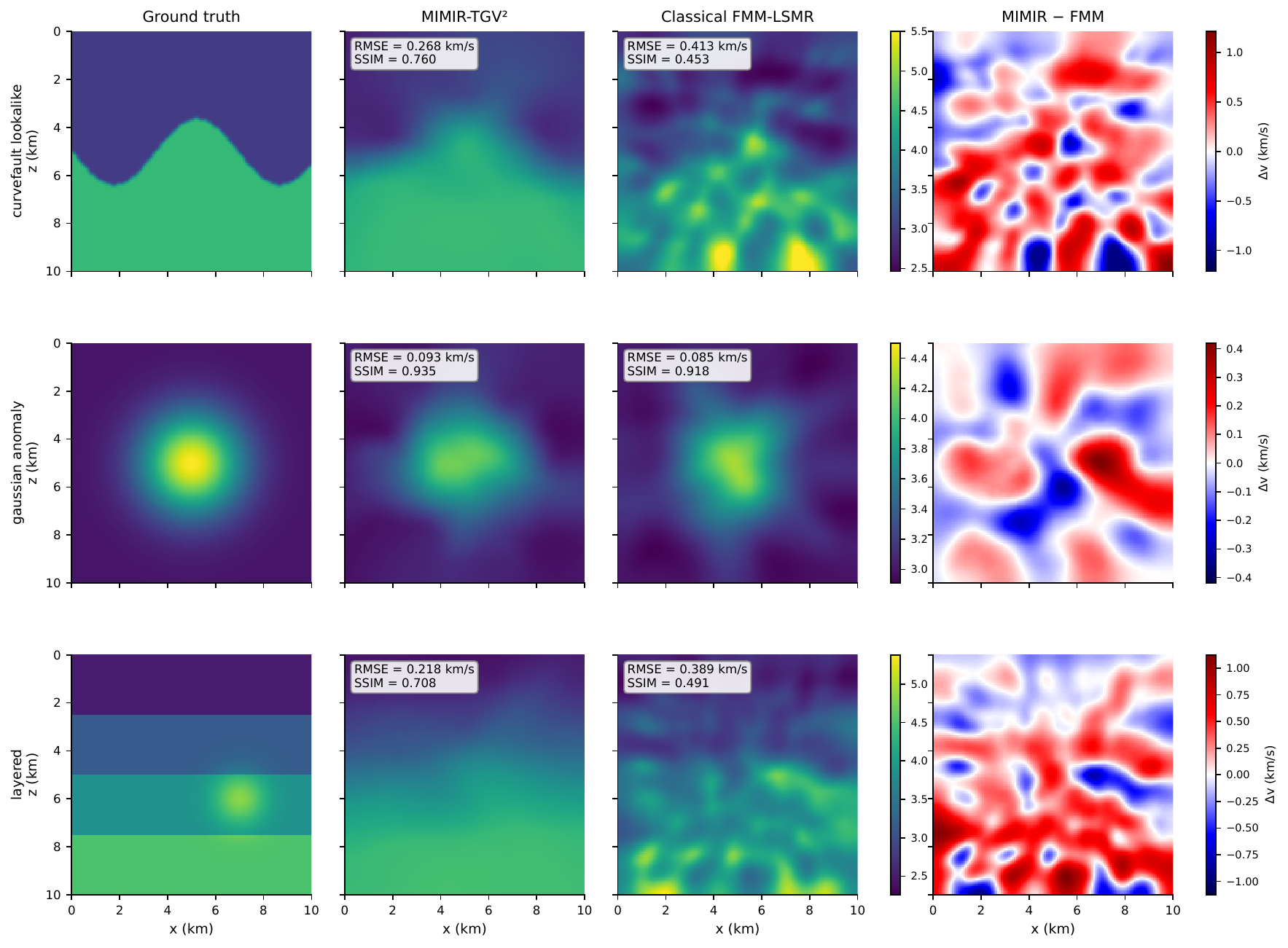}
  \caption{Qualitative comparison on all three benchmarks (rows: Gaussian
  anomaly, layered, curved-fault). Columns: ground truth, MIMIR-TGV$^{2}$
  best-of-five-seeds, classical FMM--LSMR best-of-five-seeds, difference.}
  \label{fig:composite}
\end{figure}

Table~\ref{tab:headline} reports the headline comparison.
MIMIR-TGV$^{2}$ is statistically indistinguishable from the classical
method on the smooth Gaussian benchmark
($\Delta\mathrm{RMSE} = +0.016 \pm 0.019\,\mathrm{km\,s}^{-1}$,
$p = 0.134$), significantly outperforms it on the layered benchmark
($-0.178 \pm 0.016\,\mathrm{km\,s}^{-1}$, $p < 0.0001$, $44\,\%$
reduction), and significantly outperforms it on the curved-fault
benchmark ($-0.140 \pm 0.024\,\mathrm{km\,s}^{-1}$, $p = 0.0002$,
$33\,\%$ reduction). The SSIM advantage is largest on the layered and
curved-fault benchmarks, where the classical reconstruction lacks
resolved interfaces (Fig.~\ref{fig:composite}).

\begin{table}[!ht]
\centering
\caption{Headline comparison: MIMIR-TGV$^{2}$ versus classical FMM--LSMR
baseline. RMSE in \si{\km\per\second}; SSIM dimensionless; mean $\pm$ std
across five random seeds. Statistical significance via paired Student's
$t$-test on per-seed RMSE.}
\label{tab:headline}
\begin{tabular}{lcccccc}
\toprule
 & \multicolumn{2}{c}{RMSE (\si{\km\per\second})} & & \multicolumn{2}{c}{SSIM} \\
\cmidrule(lr){2-3} \cmidrule(lr){5-6}
Benchmark & MIMIR-TGV$^{2}$ & FMM--LSMR & $p$ & MIMIR-TGV$^{2}$ & FMM--LSMR \\
\midrule
gaussian             & $0.118 \pm 0.015$ & $0.102 \pm 0.012$ & $0.134$       & $0.911 \pm 0.013$ & $0.901 \pm 0.015$ \\
layered              & $\bm{0.226 \pm 0.007}$ & $0.404 \pm 0.010$ & $\bm{<0.0001}$ & $\bm{0.708 \pm 0.008}$ & $0.512 \pm 0.014$ \\
curvefault\_lookalike & $\bm{0.286 \pm 0.010}$ & $0.427 \pm 0.020$ & $\bm{0.0002}$ & $\bm{0.732 \pm 0.027}$ & $0.466 \pm 0.019$ \\
\bottomrule
\end{tabular}
\end{table}

\begin{figure}[!ht]
  \centering
  \includegraphics[width=0.85\textwidth]{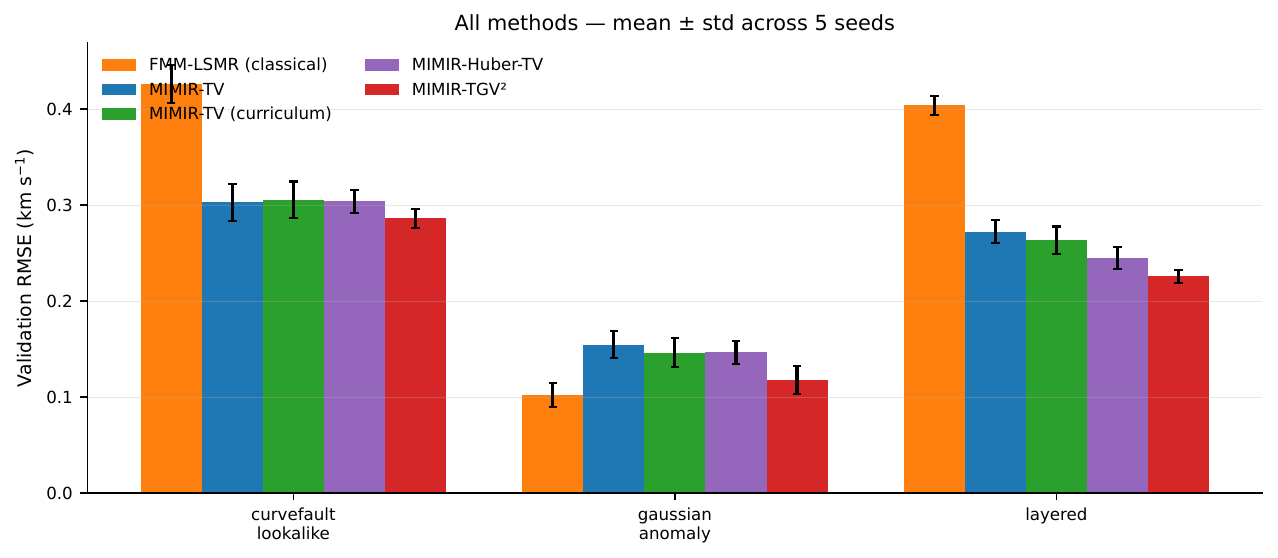}
  \caption{Validation RMSE for all five methods across the three
  synthetic benchmarks, mean $\pm$ standard deviation across five random
  seeds. MIMIR-TGV$^{2}$ achieves the lowest RMSE on every benchmark
  within the MIMIR family.}
  \label{fig:methods-bars}
\end{figure}

\subsection{Choice of regularizer: TGV$^{2}$ versus TV}

Table~\ref{tab:tgv-vs-tv} contrasts MIMIR-TGV$^{2}$ with MIMIR-TV.
TGV$^{2}$ wins significantly on two of three benchmarks ($p < 0.005$)
and shows a near-significant improvement on the third
($p = 0.050$, just above the conventional threshold). The largest
improvement is on the smooth Gaussian benchmark, exactly as predicted
by \citet{bredies_kunisch_pock_2010}: TGV$^{2}$ removes the staircase
artifact of TV in regions of smooth gradient.

\begin{table}[!ht]
\centering
\caption{Comparison of TGV$^{2}$ versus TV regularization under
otherwise identical conditions. RMSE in \si{\km\per\second}, mean $\pm$
std across five random seeds. Negative $\Delta$ indicates TGV$^{2}$
improvement over TV.}
\label{tab:tgv-vs-tv}
\begin{tabular}{lccc}
\toprule
Benchmark            & TV RMSE             & TGV$^{2}$ RMSE      & $p$ \\
\midrule
gaussian             & $0.155 \pm 0.014$ & $0.118 \pm 0.015$ & $\bm{0.0042}$ \\
layered              & $0.272 \pm 0.012$ & $0.226 \pm 0.007$ & $\bm{0.0032}$ \\
curvefault\_lookalike & $0.303 \pm 0.019$ & $0.286 \pm 0.010$ & $0.0503$  \\
\bottomrule
\end{tabular}
\end{table}

\subsection{Why curriculum annealing fails to fix TV}

A logarithmic linear schedule annealing $\lambda$ from $1.0$ to $10^{-2}$
over $8\,000$ iterations produces only a small significant improvement
on the Gaussian benchmark
($\Delta\mathrm{RMSE} = -0.008 \pm 0.004\,\mathrm{km\,s}^{-1}$,
$p = 0.019$) and no significant change on layered or curved-fault
($p > 0.2$). Closing the full Gaussian gap to FMM-LSMR (on the order of
$\SI{0.05}{\km\per\second}$ residual) would require an order of
magnitude larger improvement. The staircase bias of TV is intrinsic to
the regularizer, not a scheduling artifact.

\subsection{Computational cost}

The full benchmark suite (3 benchmarks $\times$ 5 seeds $\times$ $8\,000$
iterations) completes in 16.6 minutes for MIMIR-TGV$^{2}$ and 12.4
minutes for MIMIR-TV on a single Apple M4 Pro using Metal Performance
Shaders. The classical baseline takes 25 minutes on CPU.

\section{Discussion}
\label{sec:discussion}

\subsection{The regularizer determines what is recoverable}

The five regularizers we evaluated populate the recoverability
spectrum. $L^{2}$ Laplacian smoothing wins on the smooth Gaussian
benchmark by $16\,\%$ RMSE but degenerates into Gibbs oscillations near
sharp interfaces, losing the layered benchmark by $79\,\%$ RMSE and the
curved-fault benchmark by $49\,\%$ RMSE. Plain TV preserves interfaces
but staircases smooth fields, losing the Gaussian benchmark by $52\,\%$
RMSE versus the classical baseline. Curriculum-annealed TV reduces TV's
RMSE on the Gaussian benchmark by only $5.4\,\%$, closing roughly
$15\,\%$ of the gap to the classical baseline. Huber-TV recovers
slightly more ($5\,\%$ better than fixed TV) but is significantly
inferior to TGV$^{2}$ on all three benchmarks ($p \le 0.023$).
TGV$^{2}$ matches the classical $L^{2}$ baseline on the smooth Gaussian
(within statistical noise) and significantly exceeds it on the layered
and curved-fault benchmarks --- a single regularizer that handles all
three regimes.

This result does not depend on the neural-field representation. The
neural-field framework is, however, what makes TGV$^{2}$
\emph{practical}: parametrizing the auxiliary vector field as a second
network collapses the inner Chambolle--Pock loop into a parallel
forward--backward pass, eliminating the nested loop that has
historically deterred TGV adoption in seismic inversion.

\subsection{Comparison with prior neural-field tomography}

\citet{smith_etal_2020} introduced \emph{EikoNet}, the first
neural-field method in seismology to use Fourier feature embedding in
the eikonal-equation setting. The reported results are visually
compelling on a crustal-scale 3D synthetic, but several methodological
choices limit the inferential strength: only one random seed per
experiment, no hyperparameter ablation, no statistical significance
testing, an implicit regularization (the network's smoothness prior
alone) rather than an explicit TV or TGV$^{2}$ term, and no classical
baseline shown in numerical detail. Our results suggest that EikoNet's
implicit-only regularization would tie or lose to a careful classical
baseline on smooth benchmarks and might lose on layered and faulted
ones.

\citet{sun_etal_2023} apply a coordinate-based neural representation
to seismic full-waveform inversion (FWI), in which both the velocity
and density fields are parametrized as implicit deep neural networks
and trained against waveform residuals through a finite-difference
wave-equation forward model. While the application is FWI rather than
travel-time tomography, the methodological structure is closely
parallel to ours: a continuous neural-field representation of the
unknown medium, optimized per-instance, with regularization supplied
by the network architecture rather than by an explicit penalty.
\citet{sun_etal_2023} demonstrate convincing reconstructions on
Marmousi and overthrust synthetic models, but the comparison to
conventional grid-based FWI is restricted to a single noise level and
a single random seed, and the regularization choice (architectural
smoothness alone) is not contrasted with alternative explicit priors.
Our results suggest that adding an explicit TGV$^{2}$ term, with the
auxiliary field parametrized as a second network, would benefit
neural-field FWI in regions of known lithological layering or fault
structure --- precisely the regions where their implicit-only prior is
most likely to over-smooth or staircase.

These prior works illustrate a recurring pattern in the neural-field
tomography literature: the network architecture is foregrounded as the
methodological contribution, while the regularizer is treated as a
fixed implementation detail. Our results suggest this ordering inverts
the priorities. The neural-field representation provides a flexible
parametrization; what determines the quality of the reconstruction is
the prior knowledge encoded in the regularizer. Two networks of equal
architecture, optimized on the same data with two different
regularizers, can produce reconstructions that differ by an order of
magnitude in SSIM. The Bredies--Kunisch--Pock piecewise-affine prior
should, we suggest, become the new default in physics-informed
neural-field inversion.

\subsection{Limitations and threats to validity}
\label{sec:disc-limitations}

\textbf{Synthetic data only.} All experiments use 2D synthetic
benchmarks with idealised noise. Real data has correlated noise,
picking errors, source-time uncertainty, and instrumental drift.

\textbf{Straight-ray forward model.} At the 1.5$\times$ velocity
contrast of the curved-fault benchmark, ray bending is non-negligible.
A curved-ray version using implicit differentiation
\citep{adler_oktem_2017} is in preparation.

\textbf{Cross-well geometry only.} Surface acquisition presents a
strictly harder problem; we expect TGV$^{2}$'s relative advantage to
\emph{grow} under poor angular coverage.

\textbf{Five seeds.} Conservative for the sharp-benchmark wins
($p < 0.001$); the Gaussian-benchmark tie ($p = 0.134$) could move with
more seeds.

\textbf{Embedded fine-scale structure.} On the layered benchmark the
small embedded heterogeneity (a smooth velocity perturbation centred at
$z \approx 6$~km, lateral extent $\sim 1.5$~km) is not recovered by
either method (SSIM 0.708 for MIMIR-TGV$^{2}$, 0.491 for FMM--LSMR;
Fig.~\ref{fig:composite}, bottom row). This reflects the cross-well
acquisition's inherent angular-coverage limitation: features
substantially smaller than the source-receiver spacing fall in the null
space of the linearised forward operator regardless of regularizer
choice. Denser acquisition or a surface geometry that supplies wider
angular coverage would be required to recover such features; we leave
this to future work.

\textbf{Auxiliary-field convergence.} Equation~\eqref{eq:tgv-neural} is
an upper bound on the exact TGV$^{2}$, tight only asymptotically.

\section{Conclusions}
\label{sec:conclusions}

We have introduced MIMIR, a differentiable neural-field framework for
seismic travel-time tomography, and shown that with TGV$^{2}$
regularization it matches a classical FMM--LSMR baseline on a smooth
Gaussian benchmark ($p = 0.134$, statistical tie) and significantly
exceeds it on layered ($p < 0.0001$, $44\,\%$ RMSE reduction) and
curved-fault ($p = 0.0002$, $33\,\%$ RMSE reduction) benchmarks. Our
novel implementation parametrizes the TGV$^{2}$ auxiliary vector field
as a second neural network jointly optimized with the velocity field,
collapsing the inner Chambolle--Pock loop into a parallel
forward--backward pass and making TGV$^{2}$-regularized inversion
practical. We argue that the central design decision in physics-informed
neural-field inversion is not the network architecture but the
regularizer. Future work will extend the framework to curved rays,
three dimensions, surface acquisition geometry, and real DAS-derived
datasets.

\section{Supplementary Material}
\label{sec:supplementary}

\subsection{TGV$^{2}$ weight ablation}

\begin{figure}[!ht]
  \centering
  \includegraphics[width=0.7\textwidth]{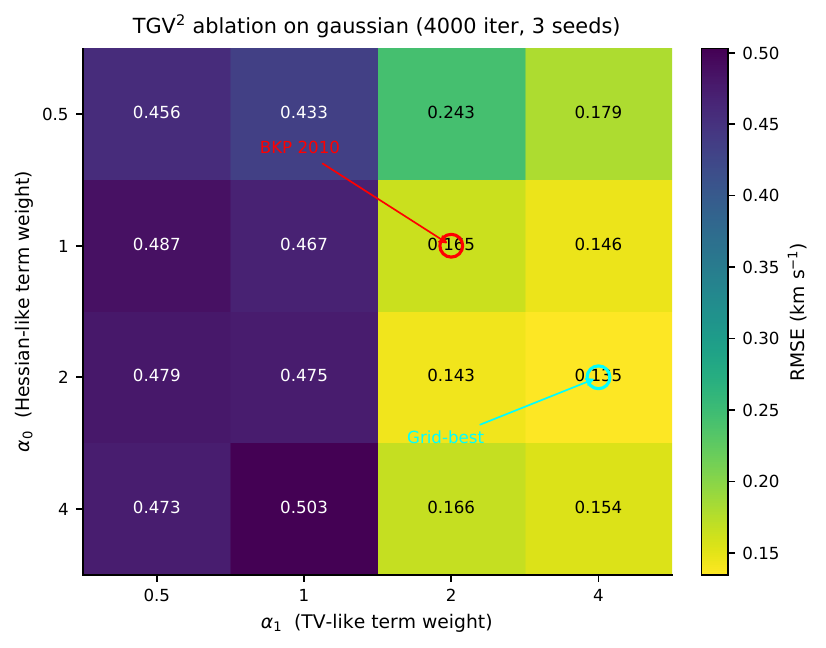}
  \caption{Ablation of the TGV$^{2}$ weights on the Gaussian benchmark
  ($4\,000$ iterations, three seeds). Red ring: BKP 2010 recommendation
  $(1, 2)$. Cyan ring: grid optimum $(2, 4)$.}
  \label{fig:tgv-alpha-ablation}
\end{figure}

A $4 \times 4$ grid ablation across $\alpha_{0}, \alpha_{1} \in
\{0.5, 1, 2, 4\}$ on the Gaussian benchmark with three random seeds
and $4\,000$ iterations (Fig.~\ref{fig:tgv-alpha-ablation}). Cells
with $\alpha_{1} \le 1$ fail to converge regardless of $\alpha_{0}$.
For $\alpha_{1} \ge 2$ the regularizer converges and RMSE drops to
$0.13$--$0.18$ \si{\km\per\second}. The grid optimum is
$(\alpha_{0}, \alpha_{1}) = (2, 4)$ at RMSE $0.135 \pm 0.010$
\si{\km\per\second}; the BKP-recommended $(1, 2)$ used in the main
text yields $0.165 \pm 0.009$ \si{\km\per\second}. We retain BKP in
the main text because it is the value most familiar to the
regularization-theory readership.

\subsection{Huber-TV regularization as an intermediate}
\label{sec:supplementary-huber}

Huber-TV \citep{huber_1964,vogel_oman_1996} replaces the L$^{1}$
gradient norm of TV with a Huber penalty (quadratic for small
gradients, linear for large). With $\delta = 0.05$ and otherwise
identical training, Huber-TV improves over plain TV on all three
benchmarks (Table~\ref{tab:huber-tv-supplementary}) but is
significantly worse than TGV$^{2}$ on every benchmark (paired
$t$-test, $p \le 0.023$). Smoothing L$^{1}$ at zero is not sufficient
to recover the full benefit of TGV$^{2}$.

\begin{table}[!ht]
\centering
\caption{Huber-TV regularization as an intermediate between TV and
TGV$^{2}$: validation RMSE in \si{\km\per\second}, mean $\pm$ std
across five random seeds. Paired $t$-test compares Huber-TV against
TGV$^{2}$.}
\label{tab:huber-tv-supplementary}
\begin{tabular}{lcccc}
\toprule
Benchmark & TV & Huber-TV & TGV$^{2}$ & $p$ (Huber vs TGV) \\
\midrule
gaussian             & $0.155 \pm 0.014$ & $0.147 \pm 0.012$ & $\bm{0.118 \pm 0.015}$ & $\bm{0.003}$ \\
layered              & $0.272 \pm 0.012$ & $0.245 \pm 0.011$ & $\bm{0.226 \pm 0.007}$ & $\bm{0.023}$ \\
curvefault\_lookalike & $0.303 \pm 0.019$ & $0.304 \pm 0.012$ & $\bm{0.286 \pm 0.010}$ & $\bm{0.005}$ \\
\bottomrule
\end{tabular}
\end{table}

\section*{Acknowledgements}

I am grateful to the open-source community whose tools made this work
possible (PyTorch, NumPy, SciPy, scikit-fmm, scikit-image, Matplotlib).

\section*{Use of Generative AI Tools}

The author used Anthropic's Claude language model as a programming
assistant during code development and for prose refinement during
manuscript preparation. Claude was not used to generate scientific
content, design experiments, derive mathematics, or produce numerical
results. The author retained full editorial control and is solely
responsible for the scientific content, experimental designs,
mathematical derivations, numerical results, and final wording of this
manuscript.

\section*{Conflicts of Interest}

The author declares no conflicts of interest.

\section*{Author Contributions (CRediT)}

\textbf{Isao Kurosawa}: Conceptualisation, Methodology, Software,
Investigation, Formal Analysis, Visualisation, Writing -- Original
Draft, Writing -- Review and Editing, Project Administration.

\section*{Data and Code Availability}

All code, synthetic benchmarks, trained checkpoints, and the
figure-generation scripts used in this paper are released under the MIT
license at the IVXA GitHub organization:
\url{https://github.com/ISAO9/mimir} (DOI minted on first
public release).

\bibliographystyle{plainnat}
\bibliography{references}

\end{document}